\documentclass[prl,twocolumn,a4paper]{revtex4-1}

\usepackage{amsmath}
\usepackage{amsfonts}
\usepackage{amssymb}
\usepackage{amsthm}
\usepackage{mathrsfs}
\usepackage{dsfont}
\usepackage{esint}
\usepackage{graphicx}

\DeclareMathAlphabet{\mathpzc}{OT1}{pzc}{m}{it}

	\newcommand{\AsymEq}{\sim}
	\newcommand{\ApproxEq}{\approx}

	\newcommand{\Ket}[1]{\left|#1\right\rangle}
	\newcommand{\sKet}[1]{|#1\rangle}
	\newcommand{\Bra}[1]{\left\langle#1\right|}
	\newcommand{\sBra}[1]{\langle#1|}
	
	\newcommand{\sBK}[2]{\langle#1|#2\rangle}
	\newcommand{\BAK}[3]{\Bra{#1}#2\Ket{#3}} 
	\newcommand{\sBAK}[3]{\sBra{#1}#2\sKet{#3}} 
	\newcommand{\KB}[2]{\left|#1\right\rangle\!\left\langle #2\right|}
	\newcommand{\sKB}[2]{|#1\rangle\!\langle #2|}

	\newcommand{\mr}[1]{\mathrm{#1}}

	\newcommand{\br}[1]{\left( #1 \right)}
	\newcommand{\brr}[1]{\left[ #1 \right]}
	\newcommand{\brrr}[1]{\left\{ #1 \right\}}
	\newcommand{\of}[1]{\!\br{#1}}
	\newcommand{\off}[1]{\!\brr{#1}}
	
	\newcommand{\sbr}[1]{( #1 )}

	\newcommand{\sof}[1]{\!\sbr{#1}}

	\newcommand{\Sum}[2]{\sum\limits_{#1}^{#2}}

	\newcommand{\Int}[3]{\int\limits_{#1}^{#2}\mr{d}#3\,}
	
	\newcommand{\sSum}[2]{\sum_{#1}^{#2}}

	\newcommand{\Integers}{\ensuremath{\mathbb{Z}} }

	\newcommand{\D}{\mathrm{d}}

	\newcommand{\Landau}[1]{\mathpzc{O}\of{#1}}

		\newcommand{\Abs}[1]{\left\vert #1 \right\vert}
		\newcommand{\sAbs}[1]{\vert #1 \vert}

		\newcommand{\Real}[1]{\mathrm{Re}\of{#1}}

		\newcommand{\PolyLog}[2]{\mathrm{Li}_{#1}\of{#2}}
		
		\newcommand{\BesselJ}[2]{J_{#1}\of{#2}}

\usepackage{wasysym}

\newcommand{\TEO}{\hat{U}}
\newcommand{\TEOz}{\hat{\mathcal{U}}}

\begin{document}

	\title{First detected arrival of a quantum walker on an infinite line}

	\author{Felix Thiel}
	\author{Eli Barkai}
	\author{David A. Kessler}
	\affiliation{
		Department of Physics, 
		Institute of Nanotechnology and Advanced Materials, 
		Bar Ilan University, 
		Ramat-Gan 52900, Israel
	} 

	\begin{abstract}
		The first detection of a quantum particle on a graph has been shown to depend sensitively on the distance $\xi$ between detector and initial location of the particle, and on the sampling time $\tau$.
		Here we use the recently introduced quantum renewal equation to investigate the statistics of first detection on an infinite line, using a tight-binding lattice Hamiltonian with nearest-neighbor hops.
		Universal features of the first detection probability are uncovered and simple limiting cases are analyzed.
		These include the large $\xi$ limit, the small $\tau$ limit and the power law decay with attempt number of the detection probability over which quantum oscillations are superimposed.
		For large $\xi$ the first detection probability assumes a scaling form and when the sampling time is equal to the inverse of the energy band width, non-analytical behaviors arise, accompanied by a transition in the statistics. 
		The maximum total detection probability is found to occur for $\tau$ close to this transition point.
		When the initial location of the particle is far from the detection node we find that the total detection probability attains a finite value which is distance independent. 
	\end{abstract}

	\maketitle
	
	\textbf{Introduction:}
		Recent experimental advances have made it possible to measure quantum walks at the single particle level \cite{Perets2008-0,Karski2009-0,Zaehringer2010-0,Xue2015-0}.
		A related advance is the quantum first detection problem which has drawn considerable theoretical attention \cite{Bach2004-0,Krovi2006-0,Krovi2006-1,Krovi2007-0,Varbanov2008-0,Stefanak2008-0,Halliwell2009-0,Halliwell2009-1,Gruenbaum2013-0,Montero2013-0,Bourgain2014-0,Krapivsky2014-0,Dhar2015-0,Dhar2015-1,Sinkovicz2015-0,Sinkovicz2016-0,Friedman2017-0,Friedman2017-1}, as it deals with the basic issue of when the particle will first be detected in a target state.
		Originally the topic  emerged in the context of quantum search algorithms.
		Given a graph, and an Hamiltonian $\hat{H}$, the presence or absence of a particle starting on node $\sKet{x_i}$ is recorded at a node $\sKet{x_d}$ sampled with period $\tau$.
		$\hat{H}$, $\tau$ and the measurement process \cite{Cohen-Tannoudji2009-0}, define the problem, which differs markedly from the corresponding well-studied classical first-passage-time problem \cite{Schroedinger1915-0,Hughes1995-0,Redner2007-0,Metzler2014-0,Benichou2015-1}.

		Recently, a quantum renewal equation which relates the statistics of first detection times to the quantum evolution operator of the measurement-free system was derived \cite{Friedman2017-0,Friedman2017-1}.
		This equation was used investigate the statistics of the first detected return, i.e. when $\sKet{x_i} = \sKet{x_d}$.
		The present Letter focuses on the first detected arrival, $\sKet{x_i} \neq \sKet{x_d}$.
		The questions to be tackled are:
		Given $\tau$ and the tight-binding Hamiltonian on an infinite line, what are the basic properties of the first detection probability?
		More specifically: 
		Will the particle always be detected? 
		If not, then what is the optimal sampling rate for which the total detection probability is maximized? 
		The existence of an optimum is expected, since the Zeno effect \cite{Misra1977-0,Itano1990-0} suppresses detection for too frequent measurements, while a too large $\tau$ aids the escape from the detector.
		What is the general behavior of the detection probability at attempt number $n$, which we denote $F_n$? 
		What is its asymptotics for small and large $n$ and how does it depend on the initial distance $\xi = \sAbs{x_d - x_i}$ of the particle from the detector?
		Surprisingly the first detection statistics become $\xi$-independent, for large $\xi$.
		Non-trivial behavior is also found for a critical sampling time for which the first detection statistics exhibit a non-analytical behavior reminiscent of a phase transition, with a qualitative change in its properties.

	\textbf{The detection protocol:}
		We focus on lattice systems.
		The system is prepared in the position eigenstate $\sKet{x_i}$.
		Detection is performed via the operator $\sKB{x_d}{x_d}$ that projects the wave function on the detection site's eigenstate.
		We repeatedly measure whether the particle is located at $x_d$ or not until we are successful.

		In the stroboscopic approach used herein, the measurement attempts are made periodically every $\tau$ units of time until a successful detection occurs; between the attempts the system evolves unitarily.
		We use the standard von-Neumann projective measurement which implies strong and instantaneous collapse of the wave function, see \cite{Cohen-Tannoudji2009-0}.
		Such a detection protocol incorporates the influence of the measurement into the remaining unitary dynamics of the system.
		The detector produces a string of length $n$ describing detection success: ``no,no,...,no,yes''.
		The wave function collapses to zero at $x_d$ after each failed detection attempt, and is then renormalized.
		After the final, successful detection, the experiment is finished and the number of detection attempts $n$ is recorded.
		The theoretical question is then to determine the probability $F_n$ that the particle is detected at $x_d$ for the first time on the $n$-th detection attempt.

		This scheme was employed in Refs. \cite{Ambainis2001-0,Bach2004-0,Krovi2006-0,Krovi2006-1,Krovi2007-0,Montero2013-0,Gruenbaum2013-0,Bourgain2014-0,Dhar2015-0,Dhar2015-1,Sinkovicz2015-0,Sinkovicz2016-0}, and by two of the current authors in \cite{Friedman2017-0,Friedman2017-1}.
		In all of these references the protocol was stroboscopic with period $\tau$, i.e. detection is attempted at times $\tau,2\tau,3\tau$, and so on.
		It was shown in Refs. \cite{Friedman2017-0,Friedman2017-1} that $\varphi_n$, the so-called detection amplitude, with $F_n = \sAbs{\varphi_n}^2$, satisfies the following quantum renewal equation:
		\begin{equation}
			\varphi_n
			=
			\BAK{x_d}{\TEO\of{n\tau}}{x_i}
			-
			\Sum{m=1}{n-1}
			\BAK{x_d}{\TEO\of{m\tau}}{x_d}
			\varphi_{n-m}
			,
		\label{eq:QuantumRenewal}
		\end{equation}
		where $\TEO\sof{\tau} = \exp\sof{- i \tau \hat{H} / \hbar}$ and $\hat{H}$ are the system's evolution operator and Hamiltonian, respectively.
		The first term on the right-hand side is the free unitary transition from $x_i$ to $x_d$ in time $n\tau$; the second term describes all previous returns to the detection site.
		We transformed Eq.~\eqref{eq:QuantumRenewal} into an equation for the generating function $\varphi\of{z}:=\sSum{n=1}{\infty}z^n\varphi_n$.
		The original amplitudes $\varphi_n$ are recovered via Cauchy's integral formula:
		\begin{equation}
			\varphi_n
			=
			\frac{1}{2\pi i}
			\oint \frac{\D z}{z^{n+1} } 
			\varphi\of{z}
			=
			\frac{1}{2\pi i}
			\oint \frac{\D z}{z^{n+1} } 
			\frac{
				\BAK{\xi}{\TEOz\of{z}}{0}
			}{
				\BAK{0}{\TEOz\of{z}}{0}
			}
			.
		\label{eq:Cauchy}
		\end{equation}
		Here $\TEOz\of{z}:=\sSum{n=0}{\infty}z^n\TEO\of{n\tau}$ is the generating function of $\TEO\sof{n\tau}$.
		$\xi:=\sAbs{x_d-x_i}$ is the distance from the initial site to the detector; we have assumed translational invariance of the system.
		The integration contour circles the origin and has to lie inside the singularities of $\varphi\sof{z}$.

	\textbf{Tight-binding model:}
		To expose the basic properties of this problem, the rest of this letter is devoted to a simple Hamiltonian which nevertheless exhibits rich physical behaviors, the 1d nearest neighbor tight binding Hamiltonian:
		\begin{equation}
			\hat{H} 
			:=
			- \gamma
			\Sum{x\in a\Integers}{} \big{[}
				\KB{x}{x+a} + \KB{x}{x-a} 
			\big{]}.
		\label{eq:DefHam}
		\end{equation}
		The lattice constant $a$ is used to rescale positions, $\xi/a\mapsto \xi$.
		Time is rescaled with $\hbar$ and the coupling energy $\gamma$, $\gamma \tau/\hbar\mapsto\tau$.
		As usual we decompose $\sBK{x}{\psi} =: \psi_x$ into position eigenstates and use the inner product $\sBK{\psi}{\chi} = \sSum{x\in a\Integers}{} \psi_x^* \chi_x$.
		The energy eigenstates of this translationally invariant operator are labeled with the wave number $k$.
		The dispersion relation is $E\sof{k}=-2\cos\sof{ak}$ in units of $\gamma$.
		Using $E\sof{k}$, we can derive an analytical representation for $\TEO\sof{n\tau}$: 
		\begin{equation}
			\BAK{\xi}{\TEO\sof{n\tau}}{0}
			=
			\frac{1}{2\pi}
			\Int{-\pi}{\pi}{k}e^{ik\xi - in\tau E\sof{k}}
			= 
			i^{\xi}
			\BesselJ{\xi}{2n\tau}
			; 
		\label{eq:TEOBessel}
		\end{equation}
		consequently $\TEOz\sof{z}$ is a series of Bessel functions.
		To build intuition we first present some numerical results.

\begin{figure}
			\includegraphics[width=0.99\columnwidth]{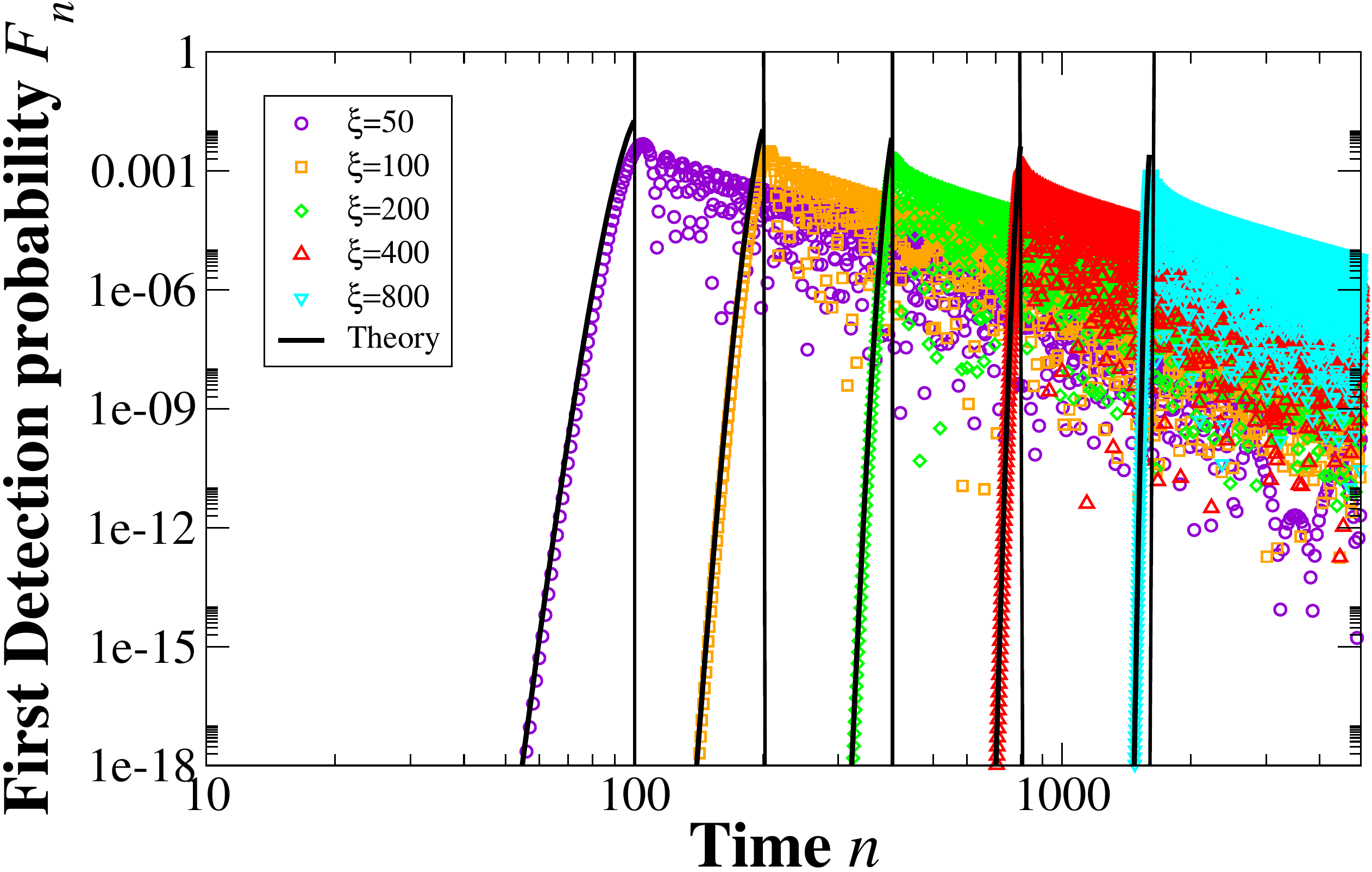}
			\caption{
				$F_n$ vs. $n$ exhibits a monotonic increase followed by oscillatory decay.
				$F_n$ (symbols) is compared to the approximation Eq.~\eqref{eq:FDPDirect}.
				Agreement is superb until the incidence time $n_{\text{inc}}$, given by Eq.~\eqref{eq:DefNInc}, that is marked with the vertical lines.
				We used $\tau=0.25$.
			}
		\label{fig:SmallN}
		\end{figure}

\begin{figure}
			\vspace{2cm}
			\includegraphics[width=0.99\columnwidth]{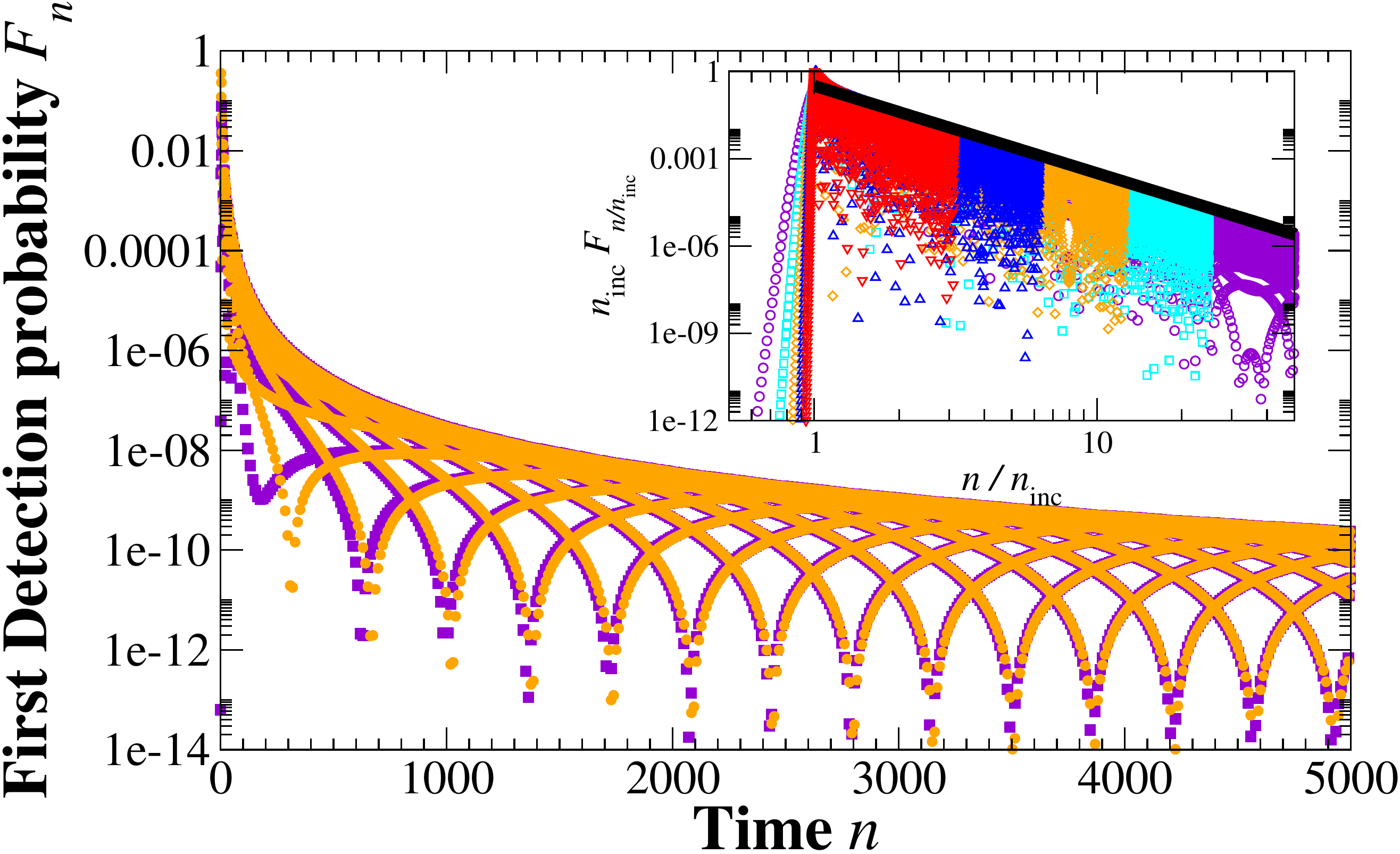}
			\caption{
				Details of the oscillatory decay.
				$F_n$ is plotted vs. $n$ for $\tau=1$ and $\xi=10$.
				The yellow circles are simulation results, the magenta squares are the approximation of Eq.~\eqref{eq:FDP}.
				A power law decay and oscillations are clearly visible in the long time limit.
				The agreement is excellent for large $n$.
				Inset: Rescaled $n_\text{inc} F_{n/n_\text{inc}}$ for $\tau=0.25$ and $\xi=50,100,200,400,800$ (colored symbols).
				The rescaling perfectly maps the envelopes to Eq.~\eqref{eq:FDPScaling}, (black line).
				In fact the curves practically overlap and all maxima move to $n/n_\text{inc}\ApproxEq1$.
				Curves with higher $\xi$ appear shorter due to rescaling.
			}
		\label{fig:LargeN}
		\end{figure}

	\textbf{Selected numerical results:}
		Eq.~\eqref{eq:QuantumRenewal} together with Eq.~\eqref{eq:TEOBessel} allow for easy numerical computation of $\varphi_n$ and $F_n$.
		We present $F_n$ for $\tau=0.25$, $\tau=1$ and different $\xi$ in Figs.~\ref{fig:SmallN} and \ref{fig:LargeN}.
		$F_n$ shows a monotonic rise to a peak for small $n$ after which it turns over to a power law decay for large $n$.
		The power law is superimposed with oscillations and has the exponent $-3$ first reported in Ref.~\cite{Bach2004-0,Friedman2017-0} for $\xi=0$.
		The position of the peak scales linearly with the distance $\xi$.
		We can align the peaks of each curve by rescaling $F_n$ and $n$ with $\xi$ and $\tau$, respectively, as shown in Fig.~\ref{fig:LargeN}'s inset.
		We now present analytical derivations of these behaviors.

	\textbf{Analytical derivation.}
		\textit{Small $n$:}
			The small $n$ behavior is dominated by the part of the wave function that directly arrives at the detection site and it is essentially unaffected by interference with returning parts.
			The lack of interference manifests itself in the monotonic increase of $F_n$, see Fig.~\ref{fig:SmallN}.
			Hence, it is a good approximation to ignore the sum in Eq.~\eqref{eq:QuantumRenewal}, as it accounts for the returning part.
			We obtain the following approximation, \footnote{Details can be found in the supplementary material (SM).}:
			\begin{equation}
				F_n 
				\AsymEq
				\Abs{J_\xi\of{2n\tau}}^2
				\brr{1 + \Landau{\tfrac{n}{\xi}}}
				=
				\frac{1}{2\pi\xi}
				\br{
					\frac{e n \tau}{\xi}
				}^{2\xi}
				\brr{1 + \Landau{\tfrac{n}{\xi}}}
				.
			\label{eq:FDPDirect}
			\end{equation}
			Thus $F_n$ grows like a large power of $n$ for large $\xi$.
			From Fig.~\ref{fig:SmallN} we see that this is essentially exact for small $n$, with significant deviations appearing where $F_n$ starts its oscillatory decay.

		\textit{Turning point:}
			The position of the turning point can be understood in terms of the maximal group velocity of the wave packet, $v_g := \max \sAbs{E'\of{k}} = 2$.
			Roughly speaking, the first component of the wave function to arrive at the destination is the one with maximal velocity.
			This happens at $n \tau = \xi / v_g$ and allows us to define the ``incidence'' time
			\begin{equation}
				n_{\text{inc}}
				=
				\frac{\xi}{v_g \tau}
				=
				\frac{1}{2}
				\frac{\xi}{\tau}
				.
			\label{eq:DefNInc}
			\end{equation}
			Note that this quantity, for a general translational invariant Hamiltonian, obviously depends on the actual energy dispersion relation $E\sof{k}$.
			The incidence time is indicated in Fig.~\ref{fig:SmallN} as well.
			It is very close to the maximum of $F_n$, as expected.

\begin{figure*}
					\includegraphics[width=0.99\columnwidth]{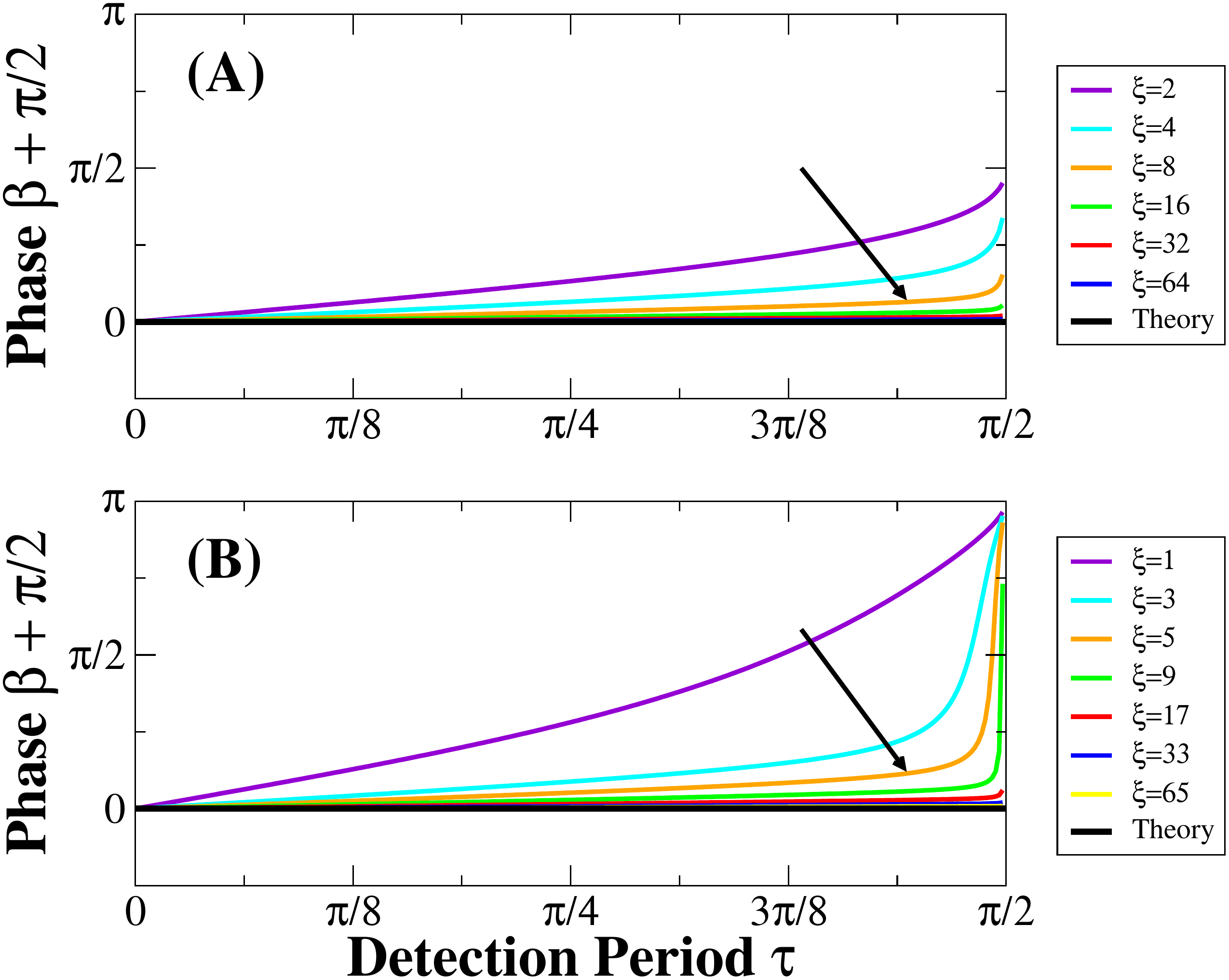}\hfill
					\includegraphics[width=0.99\columnwidth]{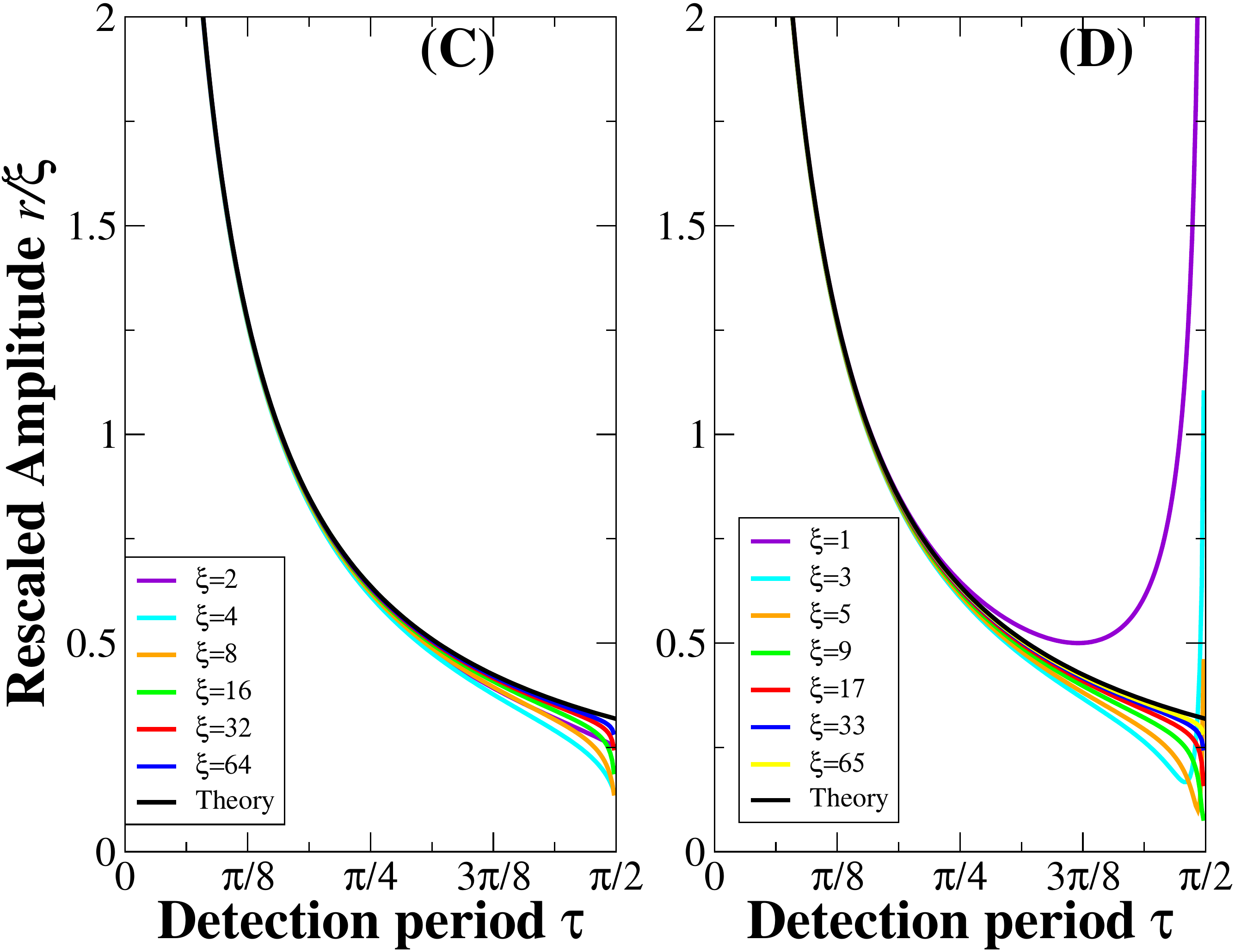} 
				\caption{
					Amplitude $r$ and phase $\beta$ of Eq.~\eqref{eq:DefR}, vs. $\tau$ and $\xi$.
					(A,B) Plot of $\beta$.
					(C,D) Plot of $r/\xi$, see Eq.~\eqref{eq:LargeXiR}.
					(A,C) Even $\xi$.
					(B,D) Odd $\xi$.
					The arrow in (A,B) denotes the direction of increasing $\xi$.
					The black line is the prediction of Eq.~\eqref{eq:LargeXiR}.
					The asymptotic values of Eq.~\eqref{eq:LargeXiR} are assumed for large $\xi$ and small $\tau$.
					In the limit $\tau\to\tau_c:=\pi/2$, $r$ diverges for odd $\xi$ and vanishes for even $\xi$.
				}
				\label{fig:Prefactor}
			\end{figure*}

		\textit{Large $n$:}
			The strategy to investigate this domain is similar to the one in Refs.~\cite{Friedman2017-0,Friedman2017-1} where the $\xi=0$ case is investigated \cite{Note1}.
			It turns out that $\varphi\sof{z}$ has two branch cuts in the complex plane along the rays with complex argument $\pm2\tau$ starting on the unit circle, see SM.
			We use those cuts to perform the complex integration of Eq.~\eqref{eq:Cauchy}.
			$\varphi\of{z}$ has $1/\sqrt{z}$-singularities at the branch points which lead to $n^{-3/2}$-asymptotics in $\varphi_n$.
			Each singularity has a complex prefactor with a phase growing linear in $n$.
			The result for $F_n$ is:
			\begin{equation}
				F_n
				\AsymEq
				\frac{4\tau}{\pi}
				r^2\of{\xi,\tau}
				\frac{
					\sAbs{\mathrm{trig}_\xi\of{2\tau n + \frac{\pi}{4} + \beta\of{\xi,\tau}}}^2
				}{n^3}
			\label{eq:FDP}
			\end{equation}
			with $r$ and $\beta$ defined by:
			\begin{equation}
				r\of{\xi,\tau} e^{i\beta\of{\xi,\tau}}
				:=
				\frac{1}{2}
				-
				\frac{i}{\pi}
				\Int{0}{\pi}{k}
				\frac{
					\sin^2\sof{\frac{\xi k}{2}}
				}{
					\tan\of{2\tau\sin^2\of{\frac{k}{2}}}
				}
				.
				\label{eq:DefR}
			\end{equation}
			The function $\mathrm{trig}_\xi\of{x} := (e^{ix}+\sbr{-1}^\xi e^{-ix})/2$ is a cosine for even $\xi$ and $i\sin\of{x}$ for odd $\xi$.
			We observe a power-law decay with exponent $-3$ superimposed with an oscillation of frequency $2\tau$.
			There is a $\xi$ and $\tau$ dependent phase $\beta$ and amplitude $r$; both are plotted in Fig.~\ref{fig:Prefactor}.
			When $\xi=0$, then $re^{i\beta}=1$ and we recover the known result \cite{Bach2004-0,Friedman2017-0,Friedman2017-1}.
			In the limit of large $\xi$ or small $\tau$ we find a remarkably simple relation to the incidence time:
			\begin{align}
				r\of{\xi,\tau} e^{i\beta\of{\xi,\tau}}
				\AsymEq &
				- i 
				\frac{\xi}{2\tau}
				= 
				- i n_\text{inc}
				.
			\label{eq:LargeXiR}
			\end{align}
			When Eq.~\eqref{eq:LargeXiR} is plugged into Eq.~\eqref{eq:FDP}, we see that the oscillation's phase becomes independent of $\xi$, but the amplitude grows quadratically in $\xi$:
			\begin{equation}
				F_n 
				\AsymEq
				\frac{1}{\pi}
				\frac{\xi^2}{\tau}
				\frac{
					\sAbs{\mathrm{trig}_\xi\of{2\tau n - \frac{\pi}{4}}}^2
				}{n^3}
				.
			\label{eq:FDPLargeXi}
			\end{equation}

			Since the amplitude $r$ becomes asymptotically proportional to the incidence time from Eq.~\eqref{eq:DefNInc}, there is a scaling form of the detection probability, that moves the maximum of $F_n$ to a fixed position, $n/n_\text{inc}\ApproxEq1$.
			The oscillation frequency cannot be rescaled as it depends on $\tau$, but we can rewrite the envelope $\bar{F}_n = 4 \tau r^2 /(\pi n^3)$ by using Eq.~\eqref{eq:LargeXiR}:
			\begin{equation}
				\bar{F}_n
				\approx
				\frac{4\tau}{\pi n_\text{inc}} \br{\frac{n_\text{inc}}{n}}^3
				,
			\label{eq:FDPScaling}
			\end{equation}
			valid for $n/n_\text{inc}>1$ and zero otherwise.
			The rescaling allows us to present the envelope as a master curve for various values of parameters, as in the inset of Fig.~\ref{fig:LargeN}, which exhibits superb agreement with theory. 

			From the definition \eqref{eq:DefR}, we find that $r\sof{\xi,\tau}$ diverges for odd $\xi$ in the limit $\tau\to\tau_c:=\pi/2$.
			At this point the two mentioned branch points of $\varphi\sof{z}$ merge into one, see SM. 
			This critical period is related to the width of the energy band and Planck's constant.
			In original units:
			\begin{equation}
				\tau_c 
				= 
				\frac{2\pi\hbar}{4\gamma}
				.
			\label{eq:CritTau}
			\end{equation}

		\textit{Small $\tau$:}
			The behavior for small $\tau$, large $n$ and fixed product $n\tau$ is given by:
			\begin{equation}
				F_n 
				\AsymEq
				\xi^2 \frac{J^2_\xi\of{2n\tau}}{n^2}
				\brr{ 1 + \Landau{\tau} }
				.
			\label{eq:FDPSmallTau}
			\end{equation}
			In particular, this approximation correctly captures the maximum of $F_n$ and the subsequently power law decay.
			Eq.~\eqref{eq:FDPSmallTau} and Eq.~\eqref{eq:FDPLargeXi} share the same asymptotics for large $n$.
			Clearly $F_n\to0$ as $\tau\to0$, which is the Zeno effect.
			The derivation and comparison to exact results can be found in the SM.

	\textbf{Total detection probability:}
		\begin{figure*}
				\includegraphics[width=0.99\columnwidth]{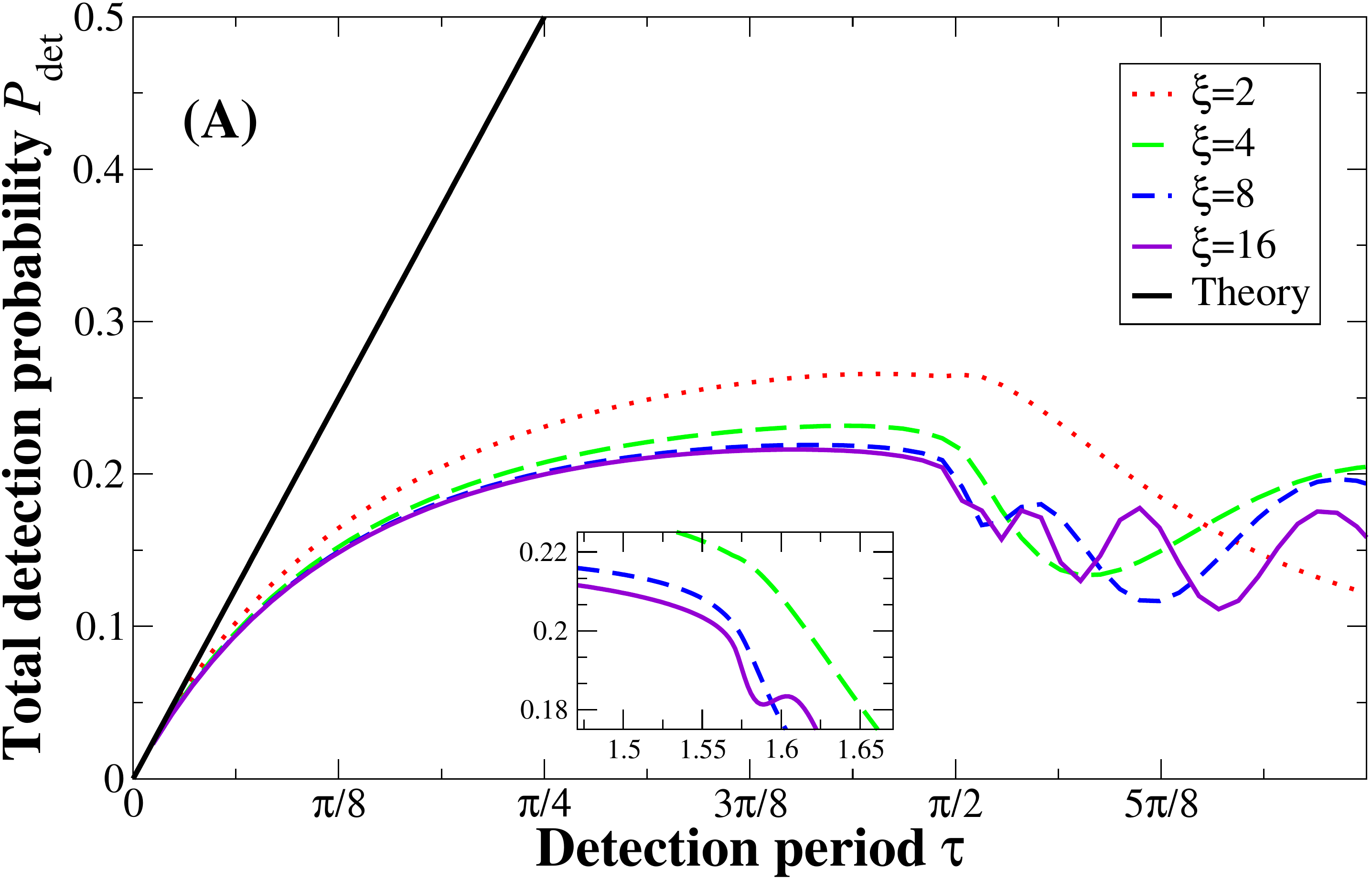}	\hfill
				\includegraphics[width=0.99\columnwidth]{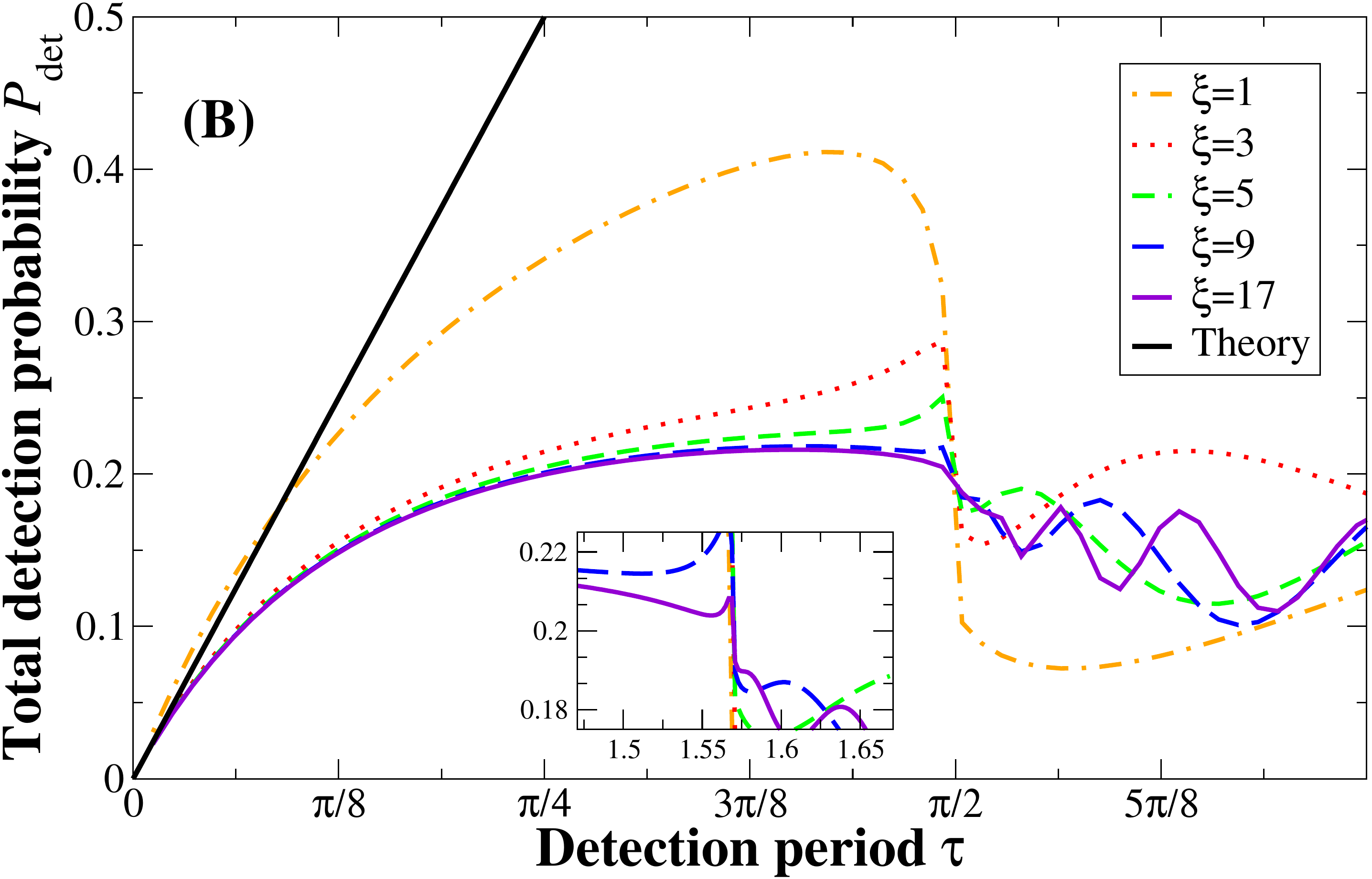}
			\caption{
				$P_\text{det}$ vs. $\tau$ for different $\xi$.
				As $\xi$ grows, $P_\text{det}$ approaches a master curve for $\tau<\tau_c=\pi/2$ ($\xi=16,17$).
				The solid black line is the prediction of Eq.~\eqref{eq:TDPSmallTau}, which only works for very small $\tau$.
				At $\tau=\pi/2$ the curve undergoes a transition from almost monotonic to oscillatory behavior.
				(B): Odd $\xi$.
				The curves are discontinuous at $\tau=\pi/2$, and the jump becomes smaller as $\xi$ increases, see Inset.
				(A): Even $\xi$. The transition is smoother here, see inset.
			}
			\label{fig:FTotal}
		\end{figure*}
		For a given $n$, $F_n$ can be small, as demonstrated in Figs.~\ref{fig:SmallN} and \ref{fig:LargeN}.
		Therefore, the question arises as to the overall probability that the particle is detected at all:
		The total probability of detection:
		\begin{equation}
			P_\text{det}
			=
			\Sum{n=1}{\infty}
			F_n
			,
		\label{eq:DefFTotal}
		\end{equation}
		is presented in Fig.~\ref{fig:FTotal} as a function of $\tau$ for several $\xi$.

		In the Zeno limit $\tau\to0$, $P_\text{det}$ approaches zero.
		It also decays for large $\tau$ (not presented).
		However, the total detection probability exhibits a very flat maximum in the vicinity of $\tau_c$, see Fig.~\ref{fig:FTotal}.
		The maximum indicates an optimal detection frequency that is related to the width of the energy band.
		The optimum balances Zeno suppression from too frequent measurements and escape from the detector arising from too sporadic detection.

		$P_\text{det}$ exhibits a transition from almost monotonic to oscillating behavior at the critical sampling time $\tau_c=\pi/2$, see Fig.~\ref{fig:FTotal}.
		For odd $\xi$, it is discontinuous at $\tau_c$ and multiples thereof.
		The jump becomes less severe as $\xi$ increases.
		As $\tau$ passes such a critical value, a new pole appears in the integrand of Eq.~\eqref{eq:DefR}.
		Since $r\of{\xi,\pi/2}$ diverges for odd $\xi$, the total detection probability exhibits a discontinuity.
		As mentioned, the critical values correspond to the points when the width of the energy band times $\tau$ is a multiple of Planck's constant.
		For this reason the critical sampling time must be universal for models with a finite energy band, and is not limited to the Hamiltonian under study.

		Very interestingly, in the limit $\xi\to\infty$, $P_\text{det}$ becomes independent of $\xi$ approaching a master curve for $\tau<\tau_c$ presented in Fig.~\ref{fig:FTotal}.
		This behavior can be understood in two ways.
		Due to the particular $\xi$-scaling of Eq.~\eqref{eq:FDPScaling}, the sum of $F_n$ is $\xi$-independent.
		Neglecting the trigonometric term and integrating $\bar{F}_n$ from $n_\text{inc}$ to infinity yields the constant $2\tau/\pi$. 
		Alternatively, one argues that the main contribution to $P_\text{det}$ comes from the peak of $F_n$ that is described by Eq.~\eqref{eq:FDPSmallTau}.
		Integrating Eq.~\eqref{eq:FDPSmallTau} over $n$ yields
		\begin{equation}
			P_\text{det}	
			\ApproxEq
			\frac{2\tau\xi^2}{\pi \sbr{ \xi^2- \frac{1}{4}}}
			\underset{\xi\to\infty}{\longrightarrow}
			\frac{2\tau}{\pi}
			,
		\label{eq:TDPSmallTau}
		\end{equation}
		which is independent of $\xi$.
		Eq.~\eqref{eq:TDPSmallTau} is valid for very small $\tau$, see Fig.~\ref{fig:FTotal}.

		We see no simple large $\xi$-limit of $P_\text{det}$ beyond the transition point $\tau_c$.
		Based on our simulations, we conjecture that $P_\text{det}$ will converge to a master curve also in this oscillatory regime, $\tau>\tau_c$.
		Our current analysis is not valid there because the integral in Eq.~\eqref{eq:DefR} diverges.
		This is left for future work.

	\textbf{Summary, Discussion and Outlook:}
		We have investigated the infinite 1d tight-binding quantum walk under repeated stroboscopic measurements with period $\tau$.
		We computed the probability of first detected arrival at attempt $n$, and its asymptotic limits for large and small $n$, and for small $\tau$.
		The distribution is highly peaked around $n_\text{inc}$,when the fastest wave packet reaches the detector.
		For large $n$, $F_n$ is given by Eqs.~\eqref{eq:FDP} and \eqref{eq:DefR}; it features strong oscillations and a power law decay with exponent $-3$.
		The frequency of $F_n$ depends only on $\tau$; its amplitude and phase depends on $\tau$ as well as on $\xi$.
		Several physical effects are found at the critical detection period $\tau_c=\pi/2$, defined by the width of the energy band:
		The amplitude $r$ diverges and $P_\text{det}$ is discontinuous for odd $\xi$.
		$P_\text{det}$ assumes its optimal value close to $\tau_c$.
		Furthermore it exhibits a transition from monotonic to oscillatory behavior at this point.
		The total probability of detection is asymptotically independent of $\xi$; this is a consequence of the surprising proportionality between the amplitude and the incidence time.
		The $\xi$-(in-)dependence of $re^{i\beta}$ and $P_\text{det}$, as well as the scaling form Eq.~\eqref{eq:FDPScaling}, are an encouraging facet of the arrival problem:
		The initial distance from the detector has only marginal impact on the first detection statistics.
		Further study will reveal whether this is an universal feature.

		Refs.~\cite{Gruenbaum2013-0,Bourgain2014-0} prove rigorously for $x_d = x_i$ that $P_\text{det} = 1$ whenever $\TEO\sof{\tau}$'s spectrum has no continuous part.
		Our system has a purely continuous spectrum and consequently we find $P_\text{det} < 1$, albeit we consider $x_d \ne x_i$.
		A spectral characterization for the continuous-spectrum systems is possible \footnote{F. Thiel, D. A. Kessler, and E. Barkai: In preparation}. 
		Such a complete spectral survey of the quantum first detection problem will elevate the theory to the standard of the classical random walk.

	\acknowledgements
		The authors acknowledge support of Israeli Science Foundation (ISF).
		FT is sustained by Deutsche Forschungsgemeinschaft (DFG) under grant no. TH2192/1-1.

	\appendix
	\section{Small $n$ limit}
		In this section we derive Eq.~(5) of the main text.
		Note that dropping the sum in the quantum renewal equation is equivalent to omitting the denominator in the generating function of $\varphi_n$.
		For $\xi>0$, the generating function $\varphi\sof{z}$ of the detection amplitudes is given by:
		\begin{equation*}
			\varphi\sof{z} 
			= 
			\frac{\sBAK{\xi}{\TEOz\sof{z}}{0}}{\sBAK{0}{\TEOz\sof{z}}{0}} 
			= 
			\sBAK{\xi}{\TEOz\sof{z}}{0}
			\brr{
				1 - \frac{\sBAK{0}{\TEOz\sof{z}}{0} - 1}{\sBAK{0}{\TEOz\sof{z}}{0}}
			}
		\end{equation*}
		The first term is the direct approximation giving the Bessel function expression:
		\begin{equation}
			\sBAK{\xi}{\TEOz\sof{z}}{0}
			:=
			i^\xi
			\Sum{n=0}{\infty} z^n \BesselJ{\xi}{2n\tau}
			.
		\label{eq:BesselExpression}
		\end{equation}
		The second term is the correction term $\epsilon_n$ that we will estimate in the following.
		Note that the remaining fraction can be identified with $\varphi\sof{z}$ for $\xi=0$.
		(We denote the original sequence here with $\varphi_n^{(0)}$.)
		As products in $z$-domain transform to convolutions in the original domain, we obtain:
		\begin{equation*}
			\epsilon_n
			:=
			\Sum{m=1}{n-1} 
			\BesselJ{\xi}{2m\tau} 
			\varphi_{n-m}^{(0)}
			.
		\end{equation*}
		For small enough $n$, the Bessel function can be expanded as $(m\tau)^\xi/\xi!$.
		Additionally, $\varphi_n^{(0)}$ is bounded by its maximum $\varphi_n^{(0)} \le c := \sup_n \varphi^{(0)}_n$.
		Using these estimates, and approximating the remaining sum over $m^\xi$ with an integral from $m=0$ to $m=n$, we find:
		\begin{align*}
			\sAbs{\epsilon_n}
			\le &
			c
			\Sum{m=1}{n-1} \sAbs{\BesselJ{\xi}{2m\tau}}
			\approx
			c
			\Int{0}{n}{m} \frac{\sbr{m\tau}^\xi}{\xi!}
			\\ = &
			\frac{\sbr{n\tau}^\xi}{\xi!} \brr{ c \frac{n}{\xi+1} }
			= \BesselJ{\xi}{2n\tau} \Landau{\frac{n}{\xi}}
		\end{align*}
		In the last line, we again identified the Bessel function of $\xi$-th order.
		Putting the direct approximation and this error estimate together, we obtain Eq.~(5) of the main text.

	\section{Large $n$ limit}
		In this section, we derive Eq.~(7) of the main text.
		Starting from Eq.~\eqref{eq:BesselExpression}, we use the asymptotic formula of the Bessel function for large arguments: $J_\xi(x) \approx \sqrt{2/(\pi x)} \cos\of{x - \xi\pi/2 - \pi/4}$.
		Asymptotically $\TEOz\of{z}$ is related to the polylogarithm function, given by:
		\begin{equation*}
			\PolyLog{\nu}{z}
			= 
			\Sum{k=1}{\infty} 
			\frac{z^k}{k^{\nu}}
			.
		\end{equation*}
		One obtains:
		\begin{align*}
			\\ &
			\BAK{\xi}{\TEOz\of{z}}{0}
			\AsymEq
			\label{eq:AsymUXi}
			\frac{i^\xi}{\sqrt{\tau\pi}}
			\Sum{j=1}{\infty} 
			\frac{z^j \cos\sof{2\tau j - \frac{\xi\pi}{2} - \frac{\pi}{4}}}{\sqrt{j}}
			\\ = & \nonumber
			\frac{1}{\sqrt{4\tau\pi}}
			\brrr{
				e^{- i\frac{\pi}{4}}
				\PolyLog{\tfrac{1}{2}}{z e^{i 2\tau}}
				+
				\sbr{-1}^\xi
				e^{i\frac{\pi}{4}}
				\PolyLog{\tfrac{1}{2}}{z e^{-i 2\tau}}
			}
			.
		\end{align*}
		Plugging this expression into Eq.~(2) of the main text yields:
		\begin{equation}
			\varphi_n 
			\AsymEq
			\frac{1}{2\pi i}
			\oint \frac{\D z}{z^{n+1} } 
			\frac{
				\PolyLog{\tfrac{1}{2}}{z e^{i 2\tau}}
				+
				i
				\sbr{-1}^\xi
				\PolyLog{\tfrac{1}{2}}{z e^{-i 2\tau}}
			}{
				\sqrt{4i\tau\pi}+
				\brrr{
					\PolyLog{\tfrac{1}{2}}{z e^{i 2\tau}}
					+
					i
					\PolyLog{\tfrac{1}{2}}{z e^{-i 2\tau}}
				}
			}
			.
			\label{eq:Cauchy2}
		\end{equation}
		The contour is a circle around the origin with subunit radius.
		\begin{figure}[t]
			\includegraphics[width=0.9\columnwidth]{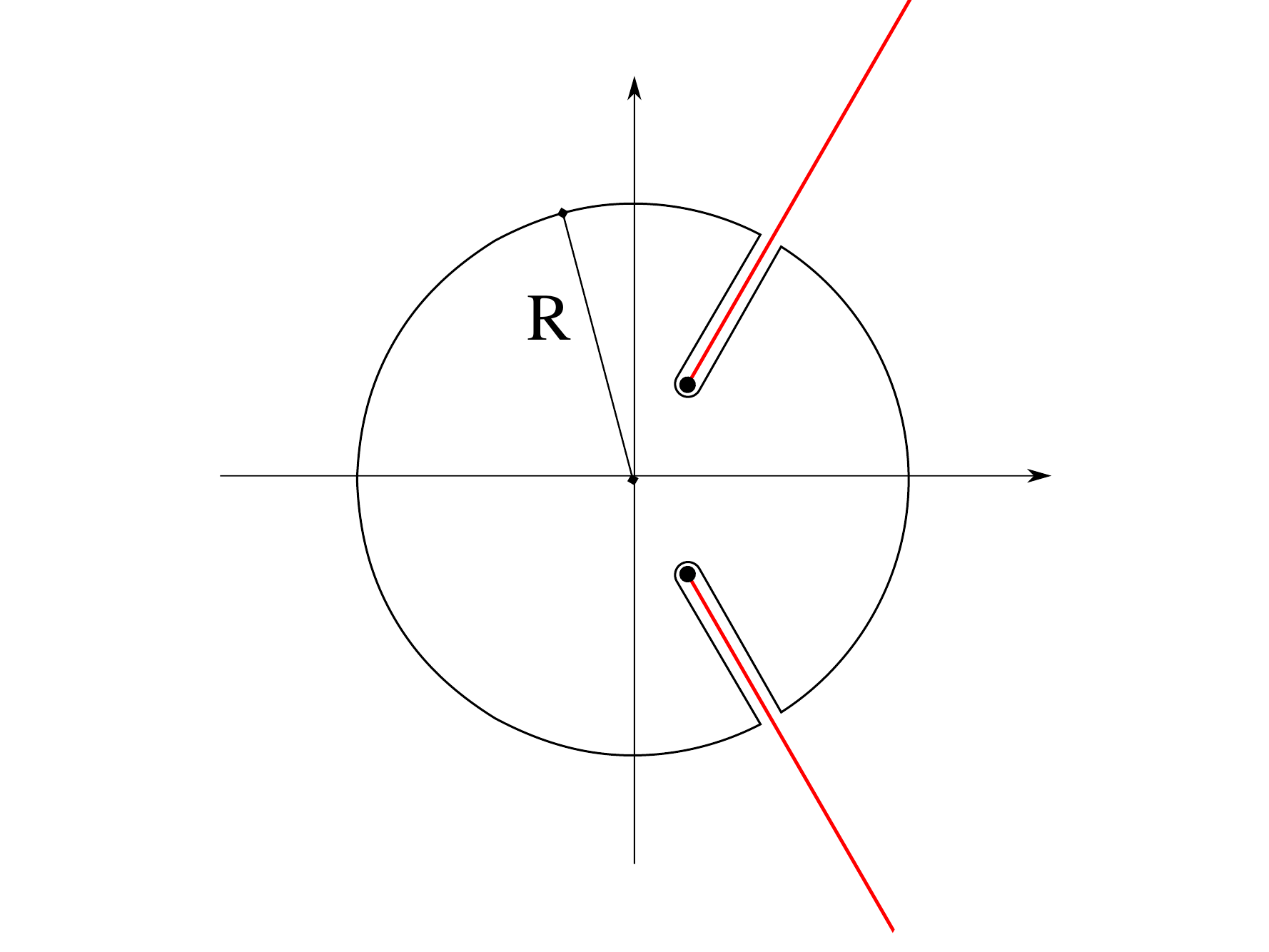}
			\caption{
				Sketch of the integration contour for Eq.~\eqref{eq:Cauchy2}.
				The integrand has two branch cuts (red lines), starting on the unit circle going to infinity  along $e^{y \pm 2i\tau}$, $y>0$.
				The radius $R$ of the contour is taken to infinity avoiding the cuts.
				The angle between the cuts is $4\tau$.
			}
			\label{fig:IntPath}
		\end{figure}

		The polylogarithm function $\PolyLog{1/2}{z}$ has a branch cut singularity starting at $z=1$ and extending along the positive real axis.
		Thus, $\sBAK{\xi}{\TEOz\sof{z}}{0}$ and also $\varphi\sof{z}$ have two branch cut singularities along the rays $z_\alpha=e^{y + \alpha 2i \tau}$, $\alpha=\pm 1$, $y \ge 0$, emerging from two points on the unit circle.
		The exception is the case when the branch cuts merge at $\tau=\pi/2$, which is the critical sampling period discussed in the main text.
		Our calculations are thus valid only for $\tau < \pi/2$.
		We deform the contour of our integral to infinite radius, returning along each branch cut, encircling it and returning to infinite radius, see Fig.~\ref{fig:IntPath}.
		Since there is no contribution from the infinite radius segments, the integral reduces to the discontinuity along each cut.
		Due to the $z^{-n-1}$, $n\gg 1$,  factor, each of these integrals along the branch cut are dominated by the small $y$ contribution.  

		For $y \ll 1$, \footnote{see functions.wolfram.com, last download May 10th 2017}
		\begin{equation*}
			\PolyLog{\frac{1}{2}}{e^y + i0^\pm}
			\AsymEq
			\pm i \sqrt{ \frac{\pi}{y} } 
			.
		\end{equation*}
		Thus
		\begin{equation*}
			\left. \PolyLog{\tfrac{1}{2}}{ze^{-2i\alpha\tau}} \right|_{z=e^{y+2i\alpha\tau}}
			\AsymEq
			\pm i\sqrt{\frac{\pi}{y}} 
		\end{equation*}
		and along the $\alpha$ branch cut
		\begin{equation}
			\left. \BAK{\xi}{\TEOz\of{z}}{0} \right|_{z=e^{y+2i\alpha\tau}}
			\AsymEq
			\pm 
			i^{1+\sbr{\alpha+1}\xi}
			\frac{e^{i\alpha\frac{\pi}{4}}}{\sqrt{4\tau y}}
			.
		\label{eq:AsymNearCircle}
		\end{equation}
		The plus (minus) sign applies, when the branch cut is approached from the counter-clockwise (clockwise) side.
		The singular behavior of $\sBAK{\xi}{\TEOz\sof{z_\alpha}}{0}$ and $\sBAK{0}{\TEOz\sof{z_\alpha}}{0}$ differs only by a factor $i^{(\alpha+1)\xi}$.
		When expanding $\varphi\sof{e^{y+2i\alpha\tau}}$ for small $y$, it is therefore convenient to reorder the integrand in Eq.~\eqref{eq:Cauchy2} and then use the small $y$ expansion Eq.~\eqref{eq:AsymNearCircle} to obtain:
		\begin{align}
			\varphi\of{e^{y+2i\alpha\tau}}
			= & \nonumber
			i^{\sbr{\alpha+1}\xi}
			-
			\frac{
				r_\alpha\of{\xi,\tau} e^{i\beta_\alpha\of{\xi,\tau}}
			}{
				\BAK{x_d}{\TEOz\of{e^{y+2i\alpha\tau}}}{x_d}
			}
			\\ \ApproxEq & 
			i^{\sbr{\alpha+1}\xi}
			\pm
			i
			e^{-i\alpha\frac{\pi}{4}}
			r_\alpha\of{\xi,\tau}e^{i\beta_\alpha\of{\xi,\tau}}
			\sqrt{4\tau y}
			\label{eq:DetAmpOnBranchCut}
			.
		\end{align}
		The numerator in the first line does not diverge when $y\to0$, but it rather assumes a complex constant.
		This is evident from the integral representation of $\TEOz\sof{z}$:
		\begin{align*}
			&
			r_\alpha\of{\xi,\tau} e^{i\beta_\alpha\of{\xi,\tau}}
			\\ := &
			\delta_{\xi,0}
			+
			i^{\sbr{\alpha+1}\xi}
			\BAK{0}{\TEOz\of{e^{i2\alpha\tau}}}{0}
			-
			\BAK{\xi}{\TEOz\of{e^{i2\alpha\tau}}}{0}
			\\ = &
			\delta_{\xi,0}
			+
			\frac{1}{2\pi}
			\Int{-\pi}{\pi}{k}
			\frac{
				i^{\sbr{\alpha+1}\xi}
				-
				e^{ik\xi}
			}{
				1-e^{2i\tau\sbr{\alpha+\cos\of{k}}}
			}
			\\ = &
			\frac{i^{\sbr{\alpha+1}\xi}+\delta_{\xi,0}}{2}
			+
			\frac{i}{2\pi}
			\Int{0}{\pi}{k}
			\frac{
				i^{\sbr{\alpha+1}\xi}
				-
				\cos\of{k\xi}
			}{
				\tan\of{\tau\sbr{ \alpha+\cos\of{k}}}
			}
			.
		\end{align*}
		In the last line, we used the identity $1/(1-e^{ix}) = (1+i \cot\sof{x/2})/2$, and the symmetry of the integral.
		Using the change of variables $k'=\pi-k$ in the last integral, we can infer the symmetry of $r_\alpha e^{i\beta_\alpha}$ under switching the sign of $\alpha$.
		Namely
		\begin{equation*}
			r_{-\alpha}\of{\xi,\tau} e^{i\beta_{-\alpha}\of{\xi,\tau}}
			=
			\sbr{-1}^\xi r_\alpha\of{\xi,\tau} e^{-i\beta_\alpha\of{\xi,\tau}}
			.
		\label{eq:}
		\end{equation*}
		This means we only need to consider one of them.
		We take $r = r_{-1}$ and $\beta = \beta_{-1}$, i.e.:
		\begin{equation}
			r\sof{\xi,\tau}e^{i\beta\sof{\xi,\tau}}
			:= 
			\frac{1+\delta_{\xi,0}}{2}
			-
			\frac{i}{\pi}
			\Int{0}{\pi}{k}
			\frac{
				\sin^2\sof{\frac{\xi k}{2}}
			}{
				\tan\of{2\tau\sin^2\of{\frac{k}{2}}}
			}
		\label{eq:DefR}
		\end{equation}
		after writing the integral in a more convenient form.
		This is Eq.~(8) of the main text.

		As the detection amplitudes on the branch cuts are given by Eq.~\eqref{eq:DetAmpOnBranchCut}, we can easily determine the discontinuity along the cut:
		\begin{equation*}
			\textrm{Disc}\off{\varphi\of{e^{y+2i\alpha\tau}}} 
			\approx 
			2 i
			e^{-i\alpha\frac{\pi}{4}}
			r_\alpha\of{\xi,\tau}e^{i\beta_\alpha\of{\xi,\tau}}
			\sqrt{4\tau y}
			.
		\end{equation*}
		Integrating along the cut gives with $z=e^{y + 2i\alpha\tau}$ and $z^{-n-1} \D z = e^{-n(2i\alpha\tau+y)} \D y$:
		\begin{align}
			\varphi_n 
			\AsymEq & \nonumber
			\Sum{\alpha}{} 
			\frac{1}{2\pi i}
			\Int{0}{\infty}{y} 
			e^{-n(2i\alpha\tau+y)}
			\textrm{Disc}\off{\varphi\of{e^{2i\alpha\tau + y}}} 
			\\ = & \nonumber 
			\Sum{\alpha}{} 
			\frac{2\sqrt{\tau}}{\pi}
			r_\alpha\of{\xi,\tau}e^{i\beta_\alpha\of{\xi,\tau}}
			e^{-2i\alpha\tau n - i\alpha\frac{\pi}{4}}
			\Int{0}{\infty}{y} \sqrt{y} e^{-ny}
			\\ = & \nonumber
			\Sum{\alpha}{} 
			\sqrt{\frac{\tau}{\pi n^3}}
			r_\alpha\of{\xi,\tau}
			e^{-2i\alpha\tau n - i\alpha\frac{\pi}{4}+i\beta_\alpha\of{\xi,\tau}}
			\\ = &
			\sqrt{\frac{4\tau}{\pi n^3}}
			r\of{\xi,\tau}
			\mathrm{trig}_\xi\of{2\tau n + \frac{\pi}{4} + \beta\of{\xi,\tau}}
			.
		\label{eq:FDA}
		\end{align}
		Here we have introduced the trigonometric function:
		\begin{equation*}
			\mathrm{trig}_\xi\of{x} 
			:= 
			\frac{e^{ix}+\sbr{-1}^\xi e^{-ix}}{2}
			=
			\left\{ \begin{aligned}
				\cos\of{x}, & \quad \xi \text{ even} \\
				i \sin\of{x}, & \quad \xi \text{ odd}
			\end{aligned} \right.
			.
		\label{eq:DefTrig}
		\end{equation*}
		$r\sof{\xi,\tau}$ and $\beta\sof{\xi,\tau}$ are defined by Eq.~\eqref{eq:DefR}.
		Squaring the modulus of Eq.~\eqref{eq:FDA} immediately leads to Eq.~(7) of the main text.

	\section{Small $\tau$ approximation}
\begin{figure}
			\includegraphics[width=0.9\columnwidth]{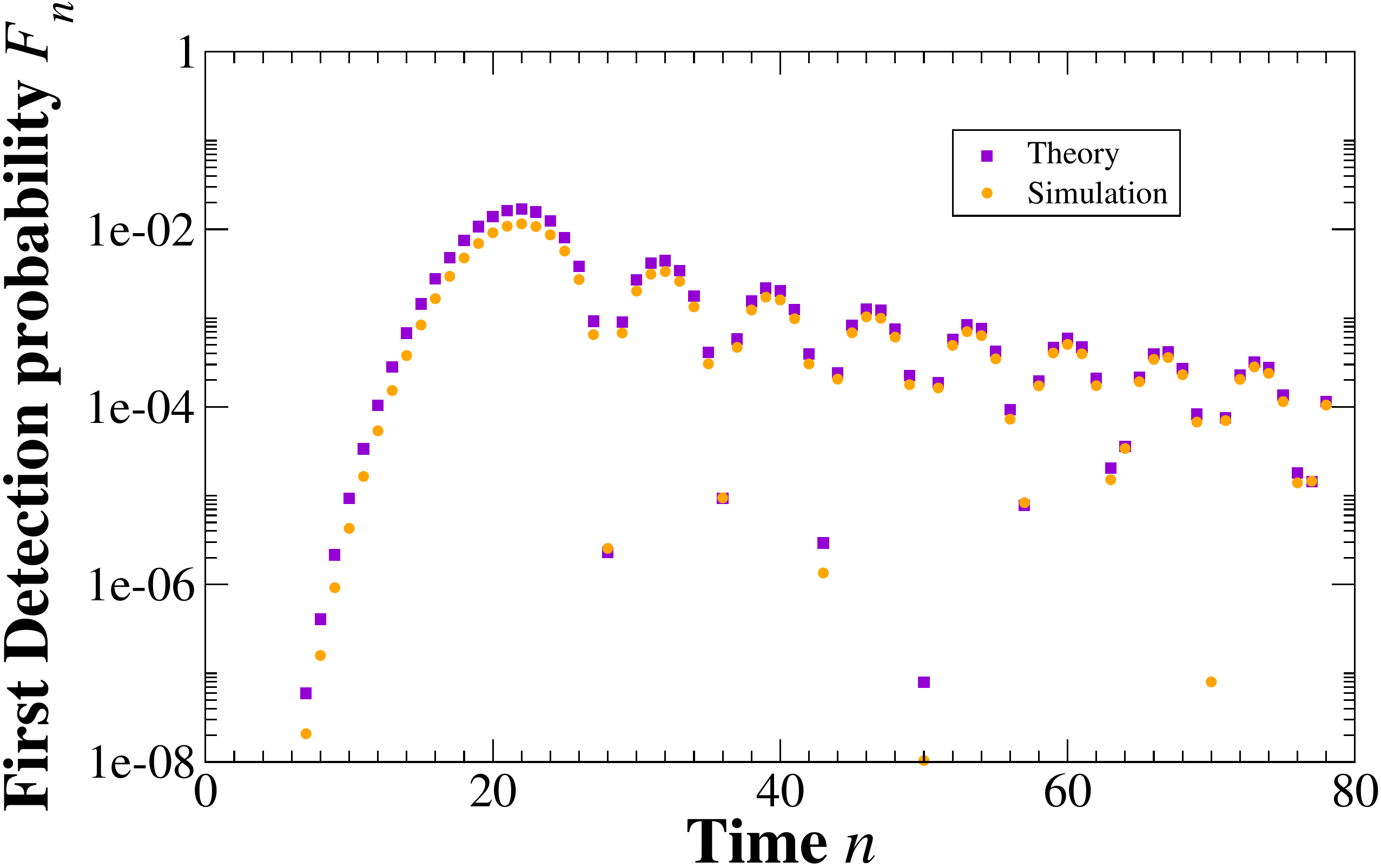}
			\caption{
				Good agreement between Eq.~(13) of the main text and simulations.
				We used $\tau=0.25$ and $\xi=10$.
				\label{fig:FSmallTau}
			}
		\end{figure}

		In this section, we derive Eq.~(13) of the main text.
		When taking the limit $\tau\to0$ it is important to rescale $n$ and $z$ simultaneously.
		This way, one does not lose important information about the first peak of $F_n$.
		We start again from Eq.~\eqref{eq:BesselExpression} and put $z = e^{-2\tau s}$ with $\Real{s} > 0$.
		This way, the argument approaches the unit circle as $\tau$ is reduced.
		When $\tau$ is very small, one can use Euler-MacLaurin formula to replace the sum with an integral:
		\begin{align*}
			& 
			2\tau \sBAK{\xi}{\TEOz\of{e^{-2n\tau s}}}{0}
			=
			2\tau i^\xi \Sum{n=0}{\infty} e^{-2\tau s n} \BesselJ{\xi}{2n\tau}
			\\ = &
			\Int{0}{\infty}{x} e^{-xs}
			\BesselJ{\xi}{x}
			+
			\Sum{l\ge0}{} \frac{2\tau B_{l+1}}{\sbr{l+1}!} \frac{\D^{l}}{\D n^{l}}
			\brr{ 
				e^{-2n\tau s}
				\BesselJ{\xi}{2n\tau}
			}_{n=0}^{\infty}
			\\ = &
			\frac{i^\xi\brr{1+\Landau{\tau^{\xi+1}}}}{\sqrt{1+s^2} \br{ s + \sqrt{1+s^2}}^\xi}
			.
		\end{align*}
		Here $B_l$ are the Bernoulli numbers. 
		To obtain the last line, the integral was identified as a tabulated Laplace transform, \cite[Eq.~103, sec. 17.13]{Gradshteyn2007-0}.
		All correction terms evaluated at $n=\infty$ vanish.
		Also, the $l$-th derivative of $\BesselJ{\xi}{2n\tau}$ vanishes at $n=0$ when $l<\xi$.
		Hence the first non-vanishing term is the one with $l=\xi$ which is of order $\tau^{\xi+1}$.
		This is the leading order in the joint limit $\tau\to0$ when $z=e^{-2\tau s}$.
		The limit $z\to1$ is equivalent to $n\to\infty$ by a Tauberian argument.
		We obtain the following expression for the generating function of the detection amplitude for $\xi>0$:
		\begin{equation*}
			\varphi\sof{e^{-2\tau s}} 
			=
			\frac{
				\BAK{\xi}{\TEOz\of{z}}{0}
			}{
				\BAK{0}{\TEOz\of{z}}{0}
			}
			=
			\frac{i^\xi\brr{ 1 + \Landau{\tau}}}{\br{s+\sqrt{1+s^2}}^\xi}
			.
		\end{equation*}
		For the inverse transform, we note that the variable transform $z=e^{-2s\tau}$ introduces an additional factor $2\tau$ in Eq.~(2) of the main text, and turns the Cauchy-integral into an inverse Laplace transform where $s$ is conjugated to $2\tau n$.
		The inverse Laplace transform is again tabulated, namely Eq.~105 of section 17.13 of \cite{Gradshteyn2007-0}.
		Consequently, we have 
		\begin{equation}
			\varphi_n
			\approx 
			2\tau
			i^\xi \xi 
			\frac{\BesselJ{\xi}{2n\tau}}{2n\tau}
			\brr{ 1 + \Landau{\tau} }
			.
		\label{eq:FDAAlt}
		\end{equation}
		The squared modulus of this expression gives Eq.~(13) of the main text.
		We have derived Eq.~\eqref{eq:FDAAlt} in a special limit where $\tau$ is small $n$ is large and the product $n\tau$ remains finite.
		A comparison between simulation and Eq.~(13) of the main text is given in Fig.~\ref{fig:FSmallTau}.

\end{document}